\documentclass[twocolumn,prl,aps,showpacs]{revtex4}

\usepackage{psfrag,graphicx}

\usepackage{dcolumn}
\usepackage{amsmath,amssymb}
\usepackage{bm}
\usepackage{pstricks}

\psset{unit=1mm,linewidth=1pt}
\def\pscr{\pscircle*}%
\def\dlad#1#2{{\multips(0,0)(7.07,0){#1}%
{\multips{45}(0,0)(3.535,-3.535){#2}{%
\psline(0,0)(5,0)(5,5)(0,5)(0,0)%
\pscr(0,0){.5}\pscr(0,5){.5}\pscr(5,0){.5}%
\pscr(5,5){.5}%
}}}}
\def\rlad#1#2{{\multips(0,0)(0,5){#2}%
{\multips(0,0)(5,0){#1}{\psline(0,0)(5,0)(5,5)(0,5)(0,0)%
\pscr(0,0){.5}\pscr(0,5){.5}\pscr(5,0){.5}%
\pscr(5,5){.5}}}}}
\def\zz#1{\multips(0,0)(10,0){#1}{%
\pscr(0,0){.5}\pscr(5,5){.5}\pscr(10,0){.5}%
\psline(0,0)(5,5)(10,0)}}

 \def\F{{\mathbb{F}}}

\def\half{\textstyle\frac{1}{2}}

\def\vec#1{{\bf{#1}}}

\begin{document}

\title[Short Title]{
Universality Classes of Diagonal Quantum Spin Ladders}

\author{
M.A. Mart\'{\i}n-Delgado$^{\star}$, J. Rodriguez-Laguna$^{\star}$ and
G. Sierra$^{\ast}$
 }
\affiliation{
$^{\star}$Departamento de
F\'{\i}sica Te\'orica I, Universidad Complutense. 28040 Madrid, Spain.
\\
$^{\ast}$Instituto de F\'{\i}sica Te\'orica, C.S.I.C.- U.A.M.,
Madrid, Spain. }

\begin{abstract}
We find the classification of diagonal spin ladders depending on a
characteristic integer $N_p$ in terms of ferrimagnetic, gapped and
critical phases. We use the finite algorithm DMRG, non-linear sigma model
and bosonization techniques to prove our results. We find stoichiometric 
contents in cuprate $CuO_2$ planes that allow for the existence of weakly
interacting diagonal ladders.
\end{abstract}


\maketitle

The determination of the ground state and low lying excited states
properties in strongly correlated systems is known to be a central
problem in condensed matter since they are responsible for their
thermodynamical properties at low temperature, their interaction with
weak external fields and generically, for the presence or not of
long-range correlations in the system.  Thus, any time a new quantum
system is brought about, their low energy properties are one of the
first issues at stake. In this work we adress this problem for the
case of diagonal quantum spin ladders with antiferromagnetic 
Heisenberg interactions:

\begin{equation}
H = J\sum_{\langle i,j\rangle\in \text{diagonal ladder}} 
\vec{S}_i\cdot \vec{S}_j,
\label{diag0}
\end{equation}

\begin{figure}[h]
\vspace{5mm}
\rput(-30,0){\psset{unit=1mm}\dlad{8}{4}
\psline{<->}(50,9)(65,-5.5)\rput[l](58,4){$N_p=4$}}
\vspace{5mm}
\end{figure}
\begin{figure}[h]
\vspace{5mm}
\rput(-30,0){\psset{unit=1mm}
\psline[linecolor=red](0,0)(60,0)
\psline[linecolor=red](5,-8.6)(65,-8.6)
\multips(0,0)(10,0){6}{\psline[linecolor=green](0,0)(5,-8.6)(10,0)}
\psline[linecolor=green](60,0)(65,-8.6)
\multips(0,0)(5,0){13}{\pscr(0,0){.5}\pscr(5,-8.6){.5}}}
\rput(36,-4){$J'$}
\rput(33,-11){$J$}
\vspace{15mm}
\end{figure}
\begin{figure}[h]
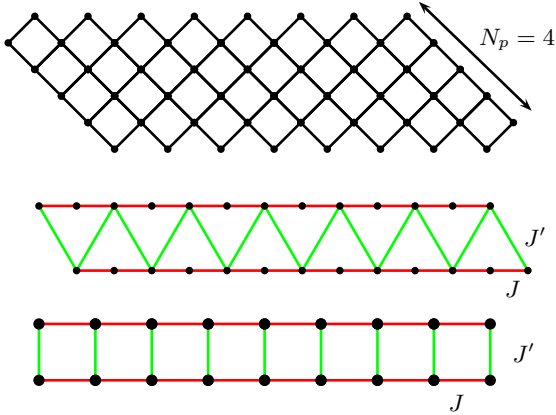

\rput(-30,0){\psset{unit=1.5mm}
\multips(0,0)(5,0){9}{\psline[linecolor=green](0,0)(0,5)}%
\multips(0,0)(0,5){2}{\psline[linecolor=red](0,0)(40,0)}%
\multips(0,0)(5,0){8}{\pscr(0,0){.5}\pscr(0,5){.5}\pscr(5,0){.5}%
\pscr(5,5){.5}}%
\rput(43,2.5){$J'$}
\rput(37,-2){$J$}}
\vspace{3mm}
\caption{a) Diagonal ladder with $N_p=4$ plaquettes. b) The $N_p=2$ diagonal
ladder as a decorated zig-zag ladder. c) Rectangular ladder with $n_l=2$ legs.}
\label{diagonalLadders}
\end{figure}

\noindent an array of sites hosting quantum spins $S=\half$ and links,
characterized by their number of plaquettes $N_p$ as depicted in
Fig.~\ref{diagonalLadders}.  On the contrary, standard square ladders
are characterized by the number of legs $n_l$ \cite{ladders}.  
The interest in this type of spin systems is manifold: 
i/ They represent an alternative
route to reaching 2D quantum physics from 1D systems \cite{diagonal}.
ii/ They appear related to the physics of stripes in cuprate materials
(e.g.$La_{1.6-x}Nd_{0.4}Sr_xCuO_4$) \cite{tranquada}.  iii/ They also
show up in numerical simulations of the $t-J$ model in planes
\cite{tJ-stripes}.  iv/ The necklace diagonal ladder is the simplest
member of the class having $N_p=1$ plaquettes. It is realized in both
organic and inorganic compounds (e.g.: metal-free poly($m$-aniline) or
poly[1,4-bis-(2,2,6,6-tetramethyl
1-4-oxy-4-piperidyl-1-oxyl)-butadiyne]) \cite{polymer-necklace}.  v/
Standard square ladders have been sinthesized in cuprate
superconductors like $Sr_{n-1}Cu_{n+1}O_{2n}, n\in\{3,5,7,\cdots\}$
\cite{ladders-cuprates}. Here we shall provide an explicit
experimental proposal to study the whole class of diagonal ladders in
cuprates.

In this paper we cast the main results of our investigations 
in a diagram of universality classes for AF diagonal
ladders depending on their number of plaquettes $N_p$:
\begin{figure}[h]
\begin{equation}
\vcenter{\hbox{Diag.}\hbox{Ladders}}
\begin{cases}
\text{Odd $N_p$}&\kern -1mm\text{Ferrimag., Gapless (F), Gapped (AF)}\\
&\\
\hbox{Even $N_p$}&\kern -1mm%
\begin{cases}N_p=2\ (mod\;4)&\text{AF, Haldane phase}\\
&\\
N_p=0\ (mod\;4)&\text{AF, Gapless}\end{cases}\\
\end{cases}
\label{universality_classes}
\end{equation}
\end{figure}

The first classification of diagonal ladders occurs depending on whether
they show ferrimagnetism or not: for $N_p$ odd they show ferrimagnetic order
while for $N_p$ even, they do not. 
The reason is as follows: although all diagonal
ladders are bipartite systems, the number of sites in each sublattice is 
different. This unbalance between the number of sites in each sublattice
is responsible for the $N_p$ odd ladders  exhibiting ferrimagnetism. 
Moreover, these  $N_p$ odd ladders present two types of excitations:
a/ ferromagnetic gapless excitations which are the 
Goldstone bosons associated to the ferromagnetic order of the ground state;
b/ antiferromagnetic gapped excitations. 

The second classification corresponds to  $N_p$ even ladders,
which are not ferrimagnetic. Their classification is not as simple as the
$N_p$ odd case since the gapless and gapped excitations do not coexists.
In fact, for $N_p=0\ (mod\;4)$ the diagonal ladders are gapless systems
with long-range interactions and critical, 
while for $N_p=2\ (mod\;4)$ they are gapped,
with finite correlation length and we can describe them as a Haldane phase.
Both types of excitations are antiferromagnetic.

Therefore, it is apparent that the classification of diagonal ladders is
richer than that of standard square ladders, which show gapless
behaviour for $n_l$ odd legs and gapped for $n_l$ even \cite{WNS}. In fact, we want
to stress that the study of diagonal ladders is much more difficult than
for square ladders. To see this, we introduce a topological deformation
of the diagonal ladders as shown in Fig.~\ref{diagonalLadders}b) for $N_p=2$.
This is achieved by flattening the outer links. The net result is a new sort of
zigzag ladder that we call {\it decorated zigzag ladder} due to the extra
sites. In the horizontal legs we put coupling constants $J$ while in the 
zigzag inter-leg links they are $J'$. This distribution
must be compared with the corresponding distribution for the square ladder
in Fig.~\ref{diagonalLadders}c) where $J'$ stands for the rung coupling 
constants. The difficulties in studying diagonal ladders can now be attributed
to the absence of a simple 
strong coupling limit $J'/J\gg 1$: when $J\gg J'$ the
ladder is equivalent to 2 weakly coupled chains (gapless point) similar to what
happens with the square ladder; when $J'\gg J$ it is again equivalent to
one weakly coupled chain (gapless point) unlike the square ladder which has
a well-defined strong coupling limit around singlet states on the vertical
rungs producing a gapped phase. The absence of a strong coupling limit
for diagonal ladders have important consequences for their study: neither it is
possible to perform a perturbation theory as in \cite{PT} nor a 
mean-field calculation based on the condensation of spin singlets on the
rungs as in \cite{GRS}. All this makes diagonal ladders a hard problem
compared to square ladders. One of the fundamental open 
questions we address is:
are there gapless phases around the limits $J\gg J'$ and $J'\gg J$?
We need hereby to resort to a series of numerical an analytical nonperturbative
methods.
\begin{figure}
\rotatebox{270}{
\includegraphics[scale=0.34]{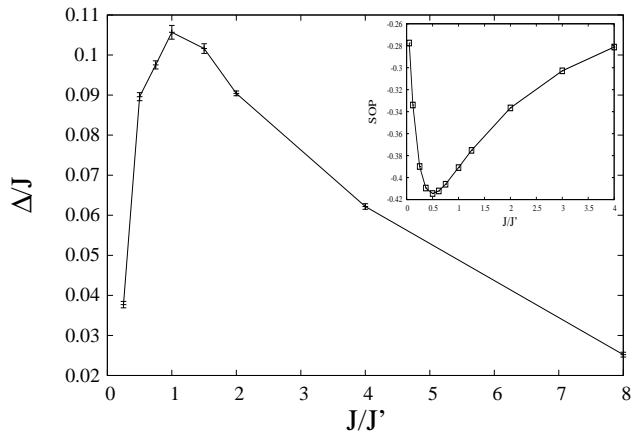}}
\caption{The gap $\Delta$ for the $N_p=2$ diagonal ladder vs. 
coupling constant. Inset: idem
for the string order paramteter.}
\label{gap}
\end{figure}

\noindent {\em DMRG}: We focus on $N_p$ even diagonal ladders and
start the DMRG calculations \cite{white1}, \cite{white2} for the
$N_p=2$ case which is equivalent to a decorated zigzag ladder
Fig.~\ref{diagonalLadders}b).  We use the finite algorithm version of
the DMRG method which is more accurate than the infinite version. We
have devised a sweeping process through the lattice adapted to the
topology of the diagonal ladders for arbitrary $N_p$ which improves
the accuracy of the algorithm: (a) all sites should be included just
once (i.e., a Hamiltonian graph); (b) ideally, all links should be
included also (i.e.,an  Eulerian graph), but since this latter is
impossible, we reduce our pretensions to (c) most of the links should
be included in the path, favouring the ones in the ``short
direction'', (d) path links which are not graph links are allowed, but
not favoured.  The maximum number of sites is $N=122$ (which
corresponds to a horizontal length of $L=30$ unit cells), the number
of states kept is $m=200$, discarded
weights of the density matrix are always below $10^{-8}$ and we
achieve a convergence of 6 digits in the energies of the eigenstates.
Moreover, we have used an independent Lanczos method to check our DMRG
results for small number of sites up to $N=28$.  In Fig.~\ref{gap} we
present our DMRG results for the gap $\Delta = E(L,S=1) - E(L,S=0)$
between the first excitated state (spin triplet) and the ground state
(spin singlet) as a function of the coupling constant ratio
$J/J'$. Each point in the plot is the result of a extrapolated fit of
the gap as a function of $1/L$ given by $\Delta(L) = a_0 + a_1 L^{-1}
+ a_2 L^{-2}$. With these results we can answer now the above open
question: we clearly see that the gap vanishes only at the extreme
cases of $J'\rightarrow 0$ and $J\rightarrow 0$.  Thus, we have a
dense gapped phase $\forall J'/J$, except at two limiting points.

\begin{figure}
\rotatebox{270}{
\includegraphics[scale=0.34]{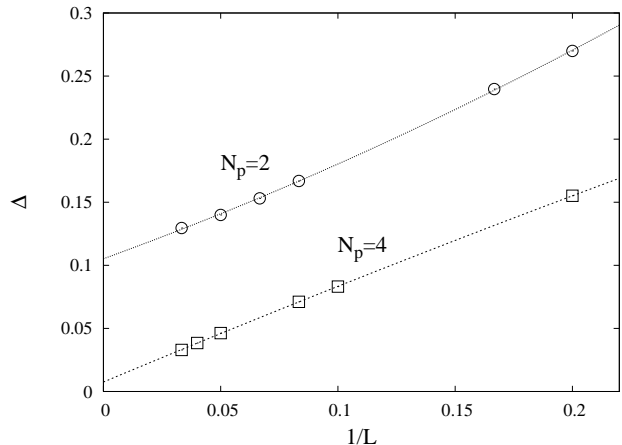}}
\caption{The gap $\Delta$ for the $N_p=4$ diagonal ladder vs. the
length $1/L$ (unit cell) for $J'=J$, and $N_p=2$ for comparison. }
\label{d2d4}
\end{figure}

The next interesting fact we find is the existence of a maximum
corresponding to the isotropic point $J=J'$. This must be compared with
the results for the standard zigzag ladder in \cite{whiteaffleck}
where the physics of the gapped phase is dominated by the
Majumdar-Gosh point (fully dimerized state) and in addition, there is
a gapless phase separated by a critical point $(J/J')_c\approx
0.24$. The reason for the standard zigzag ladder having two phases is
because it is a frustrated lattice, unlike our decorated zigzag ladder
which is bipartite and thus nonfrustrated. This makes a big difference
and is responsible for the presence of an single gapped phase in our
ladder.  Likewise, we may wonder whether our gapped phase in
Fig.~\ref{gap} is a real Haldane phase \cite{Haldane, AKLT} 
or dimerized. To clarify this
issue we have computed with DMRG the dimerized order parameter
$D^{\alpha}(L)$ for a given length as the difference between two
consecutive spin-spin correlators in the measured in the middle of the
ladder as

\begin{equation}
D^{\alpha}(L) := 
\langle \vec{S}^{\alpha}_{i-1}\cdot \vec{S}^{\alpha}_{i}\rangle_{\frac{L}{2}}
- \langle \vec{S}^{\alpha}_{i}\cdot \vec{S}^{\alpha}_{i+1}\rangle_{\frac{L}{2}}
\label{diag1}
\end{equation}
where $\alpha=u,d,z$ stands for up, down or zigzag
directions in the ladder.  For the Majumdar-Gosh state this parameter
takes a maximum value and detects its dimerization. In our $N_p=2$
ladder we find values $D^{u,d}(L=25)\approx 4\cdot 10^{-6}$,
$D^{z}(L=25)\approx 2\cdot 10^{-5}$.  Thus, our ladders are not
dimerized, a result compatible with a Haldane phase. Moreover, we have
also computed with DMRG the string order parameter (SOP) 
\cite{string} defined as
\begin{equation}
S(L) := 
\langle (\vec{S}^{u}_{1} + \vec{S}^{d}_{1})  
{\rm e}^{{\rm i}\pi \sum_{j=2}^{L-1} (\vec{S}^{u}_{j} + \vec{S}^{d}_{j})} 
(\vec{S}^{u}_{L} + \vec{S}^{d}_{L})\rangle.
\label{diag2}
\end{equation}
Non-vanishing values of (\ref{diag2}) detect the hidden order characteristic
of the Haldane phase. In Fig.~\ref{gap} (inset) we plot the extrapolated
values $S(\infty)$ and find finite values in whole range corresponding to the
gapped phase. It is remarkable that our values of the SOP are bigger than
for the square ladder, while the gap in the diagonal ladder is  more
than ten times smaller than in the square ladder.

For $N_p=4$ diagonal ladder, we have performed similar DMRG calculations
up to $N=184$ sites with $m=400$ states. Our results are shown in 
Fig.~\ref{d2d4} where we present the 
corresponding gap vs. $1/L$, and also the same for $N_p=2$ for comparison.
We find that the extrapolated value points to a negligible gap for the
$N_p=4$ ladder. This is a crucial difference w.r.t. the previous case and
it means that there is a sharp difference between the whole class of 
$N_p$ even diagonal ladders. Is it possible to give an explanation for
this difference?

\begin{figure}
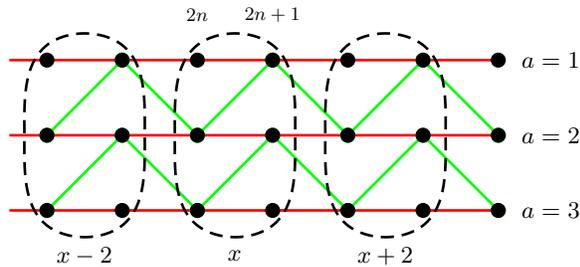

\rput(-30,-20){\psset{unit=1mm}
\psline[linecolor=red](-5,0)(60,0)
\psline[linecolor=red](-5,10)(60,10)
\psline[linecolor=red](-5,-10)(60,-10)
\multips(0,0)(0,-10){2}{\psset{unit=2mm,linewidth=1pt,linecolor=green}\zz{3}}
\multips(0,0)(10,0){7}{\pscr(0,0){1}\pscr(0,10){1}\pscr(0,-10){1}}
\multips(0,0)(20,0){3}{\psccurve[linestyle=dashed,linewidth=1pt]%
(0,12)(10,12)(13,0)(10,-12)(0,-12)(-3,0)}
\rput(5,-16){$x-2$}%
\rput(25,-16){$x$}%
\rput(45,-16){$x+2$}%
\rput(67,10){$a=1$}
\rput(67,0){$a=2$}
\rput(67,-10){$a=3$}
\rput(20,16){\scalebox{.8 .8}{$2n$}}
\rput(30,16){\scalebox{.8 .8}{$2n+1$}}}
\vspace{35mm}
\caption{Block-decomposition suitable for diagonal ladders.}
\label{blocking}
\end{figure}

\noindent {\em Non-Linear Sigma Model:}
We turn to the sigma model (NLSM) \cite{Haldane} to answer this question.
The low energy properties of a Heisenberg lattice can be mapped onto 
the effective Hamiltonian of a non-linear sigma model given by
\begin{equation}
{\cal H}=\frac{v_{\sigma}}{2}\int dx \left[ g\left(\vec{\ell} - 
\frac{\theta}{4\pi}\vec{\phi}'\right)^2
+ \frac{1}{g}\vec{\phi}'^2\right],
\label{dia3}
\end{equation}
where $\vec{\phi}$ is an $O(3)$ sigma field with the constraint
$||\vec{\phi}||^2=1$, $\vec{\ell}$ is the angular momentum, 
$v_{\sigma}$  the spin wave velocity, $g$ the counpling constant 
and $\theta$ is the topological term which is defined only mod $2\pi$.
These $(v_{\sigma},g,\theta)$ are phenomenological parameters that
will depend on the microscopic parameters like $(J,J',S)$ 
via the sigma model mapping \cite{Haldane}. The  behaviour of the system 
depends on whether the topological
constant $\theta=0 \;(\text{mod}\; 2\pi)$ for gapped or 
$\theta=\pi \;(\text{mod} \;2\pi)$ for gapless. 
This mapping can be done with a real-space blocking procedure  
\cite{sierra}. For the diagonal ladders we need to use the topological 
deformation introduced earlier in Fig.~\ref{diagonalLadders}b) in order
to bring them to the form of generalized decorated ladders with 
$n_l=\frac{N_p}{2}+1$ legs as shown also in  Fig.~\ref{blocking}, where
we also show the splitting of the original Hamiltonian (\ref{diag0})
into a block $H_B$ and interblock $H_{BB}$ parts. In order to make an
appropriate ansatz we need to take into account several subtleties
occurring in diagonal ladders: a/ the Neel state has AF character in
the horizontal direction but Ferromagnetic in the vertical direction
despite being all the couplings $J,J'$ positive (Fig.~\ref{blocking}); 
b/ the even-site spin operators are inequivalent to the odd-site spins
due to the peculiar zigzag interleg couplings (Fig.~\ref{blocking}).
With these provisos in mind, we find the correct ansatz to be: 
\begin{equation}
\begin{cases}
\vec{S}_a(2n) &= A_a^e\ell(x) + \ell_a(x) + S\left(\phi(x)+\phi_a(x)\right) \\
\vec{S}_a(2n+1)& = A_a^o\ell(x) + \ell_a(x) - S\left(\phi(x)+\phi_a(x)\right)
\end{cases}
\label{dia4}
\end{equation}

\noindent for the spin operators within each block in terms of sigma
model fields and extra massive fields $\vec{\phi}_a, \vec{\ell}_a$
needed to account for the correct number of degrees of freedom in each
block.  $A^{e,o}_a$ are amplidutes computed from spin wave theory in
terms of the microscopic parameters. To fulfill the canonical
commutation relations between the sigma model fields $\vec{\phi}$ we
need to impose the constraints $\sum_a A^e_a = \sum_a A^o_a = 1$,
$\sum_a \ell_a=0$, $\sum_a \phi_a=0$.  After a lengthy calculation we
find the result:
\begin{equation}
\theta = 2\pi S\left(\frac{N_p}{2}+1\right).
\label{diag5}
\end{equation}
\begin{table}
\begin{ruledtabular}
\begin{tabular}{|c|c|c|}
 & $N_p=2$ \text{Diagonal Ladder} & $N_p=4$ \text{Diagonal Ladder}\\
\hline
$\theta$ & $4\pi S$ & $6\pi S$ \\ \hline
$g$ & $\frac{1}{S}\left(1+\frac{J'}{J}\right)^{-1/2}$ & 
$\frac{2}{3S}\left(1+\frac{4J'}{3J}\right)^{-1/2}$\\ \hline
$v_{\sigma}$ & $2SJ\left(1+\frac{J'}{J}\right)^{1/2}$ & 
$2JS\left(1+\frac{4J'}{3J}\right)^{1/2}$\\ 
\end{tabular}
\end{ruledtabular}
\caption{NLSM data $(v_{\sigma},g,\theta)$ in $N_p$-even diagonal 
ladders.}
\label{table}
\end{table}
This is independent of $J$ and $J'$, a rather remarkable result having
in mind the differences introduced by the zigzag couplings.  We also
calculate the values of $g$ and $v_{\sigma}$ in Table~\ref{table}.
The coupling constant has very interesting information that clarifies
the issue of lack of strong coupling in diagonal ladders, namely:
\begin{equation}
\begin{cases}
g \sim (\frac{J'}{J})^{-\half} \longrightarrow 0, \; \; \;\; \;N_p \: \text{even diagonal ladders,}\\
g \sim \begin{cases}
(\frac{J'}{J})^{\half} \longrightarrow \infty,& \; n_l \;\text{even square ladders,}\\
\text{const.} & \; n_l \;\text{odd square ladders.}
\end{cases}
\end{cases}
\label{diag6}
\end{equation}

\noindent Thus, it vanishes in the strong coupling limit $J'\gg J$
meaning that the system becomes less disordered.  This is in sharp
contrast with the behaviour of square ladders.  With this NLSM
calculations we see that $N_p$ even ladders can have a very different
behaviour as shown in Fig.~\ref{universality_classes} and the DMRG
calculations in Fig.~\ref{d2d4}.

\noindent {\em Bosonization:} We can obtain additional information
about the gap formation in the $N_p=2$ diagonal ladder (decorated
zigzag) using abelian bosonization techniques. Each leg of the
decorated zigzag ladder (Fig.~\ref{diagonalLadders}b)) can be
described in the continuum limit in terms of massless Bosonic fields
$\varphi_a(x), a=1,2$ with a non-interacting Hamiltonian $H_0=
\frac{v_s}{2}\sum_{a=1,2}\int dx \left[ \Pi_a^2(x)+
(\partial_x\varphi_a(x))^2\right]$.
The corresponding SU(2) currents
$\vec{J}^{(L,R)}_{a}(x)$ and staggered field $\vec{n}_a(x)$
satisfy 
$\vec{S}_a(x) = \vec{J}^L_{a}(x) + \vec{J}^R_{a}(x) + (-1)^n\vec{n}(x)$.
For weak coupling regime $J'/J\ll 1$, we may switch on the interchain
coupling
\begin{equation}
H_{\text{int}} = J'\sum_{n}\left[ \vec{S}_1(2n+1)+\vec{S}_1(2n+3)\right]
\cdot \vec{S}_2(2n+2),
\label{}
\end{equation}
and bosonize it with the result 
$H_{\text{int}} \approx -4 J' a_0 \int dx \; \vec{n}_1(x)\cdot  \vec{n}_2(x),$
where we have kept only the strongly relevant staggered
parts and the marginal current-current terms are discarded. The nice
feature of this result is that when we compare it with the
bosonization of the standard square 2-leg
ladder $H_{\text{int}} \approx J'
a_0 \int dx \; \vec{n}_1(x)\cdot \vec{n}_2(x)$ \cite{whiteaffleck} and
the standard zig-zag ladder $H_{\text{int}} \approx J' a_0 \int
dx \; [-(\partial\vec{n}_1(x))\cdot \vec{n}_2(x)+
(\partial\vec{n}_2(x))\cdot \vec{n}_1(x)]$ we find that our decorated
zigzag ladder ($N_p=2$) produce a gap of the same type as the square
ladder (both unfrustrated) and not related to the standard zigzag
ladder which is dimerized.  Therefore, this supports
the existence of a Haldane phase in our $N_p=2$ diagonal ladder.

\noindent {\em Experimental Proposal:} Let us now turn to the
experimental realization of diagonal ladders mentioned at the beginning
based on cuprate materials. We find that it is possible to modify the
basic  $CuO_2$-stoichiometry of the cuprate planes in order to arrange
the $Cu$ sites and the $O$ sites in links of $180$ degrees with the
precise shape of a diagonal ladder, while each diagonal ladder are
weakly interacting among each other because they share links of
$Cu-O-Cu$ making $90$ degrees that are much weaker. This construction
for the Heisenberg interaction (\ref{diag0}), is based on the
superexchange mechanism among $Cu$ sites mediated by interstitial $O$
sites which is stronger when the sites $Cu-O$ are aligned in $180$
degrees.  An example of our construction for the first $N_p=1, 2$
diagonal ladders (\ref{universality_classes}) is
shown in Fig.~\ref{cuprates_series}, corresponding to a stoichiometry
of $Cu_3O_4$ and $Cu_2O_3$, respectively. 
\def\F#1{\multips(0,#1)(5,0){26}{\pscr(2.5,2.5){.8}}}%
\def\H#1{\multips(0,#1)(10,0){26}{\pscr(2.5,2.5){.8}}}%
\def\HH#1{\multips(-5,#1)(10,0){26}{\pscr(2.5,2.5){.8}}}%
\newrgbcolor{darkgreen}{0 0.5 0}

\begin{figure}[h]
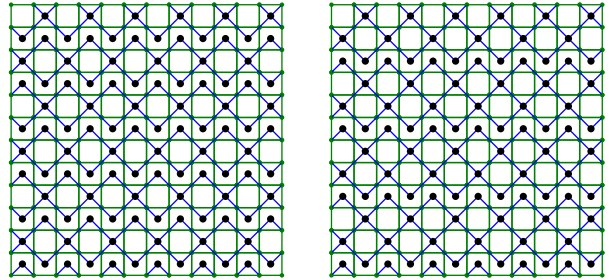

\psset{unit=.6mm,linewidth=.5pt}
\rput(-65,-58){{\psset{linecolor=darkgreen}\rlad{12}{12}}%
\psclip{\psframe[linestyle=none](0,0)(60,60)}%
\multips(-35,-5)(5,10){8}{\multips(0,0)(10,0){18}{%
\pspolygon[linecolor=blue](2.5,2.5)(7.5,7.5)(12.5,2.5)(7.5,-2.5)}}%
\multips(0,0)(0,20){3}{\F{0}\H{5}\F{10}\HH{15}%
}%
\endpsclip}%
\rput(6,-58){{\psset{linecolor=darkgreen}\rlad{12}{12}}%
\psclip{\psframe[linestyle=none](0,0)(60,60)}
\multips(-50,-10)(10,15){8}{\multips(0,0)(10,0){18}{%
\pspolygon[linecolor=blue](2.5,2.5)(7.5,7.5)(12.5,2.5)(7.5,-2.5)
\psline[linecolor=blue](7.5,7.5)(12.5,12.5)(17.5,7.5)}}%
\multips(0,0)(0,15){7}{\F{0}\H{5}\HH{10}}%
\endpsclip}
\vspace{35mm}
\caption{$N_p=1$ (left) and $N_p=2$ (right) weakly interacting 
diagonal ladders in cuprate planes.
Solid circles are Cu sites and the lattice sites are the O interstitial sites. }
\label{cuprates_series}
\end{figure}
The outcome of our general construction is summarized in the following
series of cuprate materials based on Stroncium:
\begin{equation}
\begin{cases}
Sr_{n-1}Cu_{n+1}O_{2n}:  n = 2,4,\cdots 
\longrightarrow N_p= 1, 3, \cdots ; \\
Sr_{n-1}Cu_{n+1}O_{2n}: n = 3,5,\cdots 
\longrightarrow N_p =2, 4, \dots .
\end{cases}
\label{}
\end{equation}

Interestingly enough, the series of cuprates with $n$ odd has been
studied experimentally as a set of square ladders with variable number
of legs $n_l$ \cite{ladders-cuprates}, both even and odd. What we have
found is that our diagonal ladders with $N_p$ even are allomorphic
with all the square ladders. This makes sense since we have found both
gapped and gapless behaviour in $N_p$ even diagonal ladders
(\ref{universality_classes}).  It is remarkable that each of the
series in our classification (\ref{universality_classes}) of
universality classes corresponds to a different series in the same
family of cuprate materials and it may allow for an experimental test
of those complex behaviours found in diagonal quantum spin ladders.

\noindent {\em Acknowledgments}. We thank S.R. White for fruitful discussions. 
This work is partially supported by the
DGES under contract BFM2003-05316-C02-01.

\end{document}